\documentclass[runningheads]{llncs}

\usepackage{natbib}
\usepackage{graphicx}
\usepackage{amsmath}
\usepackage{amssymb}

\begin{document}

\title{Real-Time Cardiac Cine MRI with Residual Convolutional Recurrent Neural Network}
\author{Eric Z. Chen\inst{1} \and Xiao Chen\inst{1} \and Jingyuan Lv\inst{2} \and Yuan Zheng\inst{2} \and Terrence Chen\inst{1} \and Jian Xu\inst{2} \and Shanhui Sun\inst{1}}

\institute{United Imaging Intelligence, Cambridge, MA, USA \and UIH America, Inc., Houston, TX, USA}

\authorrunning{E. Chen et al.}
\titlerunning{ }

\maketitle

\begin{abstract}
Real-time cardiac cine MRI does not require ECG gating in the data acquisition and is more useful for patients who can not hold their breaths or have abnormal heart rhythms. However, to achieve fast image acquisition, real-time cine commonly acquires highly undersampled data, which imposes a significant challenge for MRI image reconstruction. We propose a residual convolutional RNN for real-time cardiac cine reconstruction. To the best of our knowledge, this is the first work applying deep learning approach to Cartesian real-time cardiac cine reconstruction. Based on the evaluation from radiologists, our deep learning model shows superior performance than compressed sensing.

\end{abstract}



 

\section{Introduction}

Real-time cardiac cine MRI (RT-cine), compared to retro-cine, requires neither ECG gating nor breath-holding, which can be applied to a more general patient cohort and simplify the scanning process \citep{yuan2000cardiac}. To achieve a high temporal resolution ($<$50ms) and an acceptable spatial resolution ($<$2mm), highly undersampled data ($>$10x acceleration) needs to be collected for RT-cine, which imposes a significant challenge for image reconstruction. Compressed sensing (CS) based approaches have been proposed for dynamic cardiac MR (CMR) \citep{chen2014real, kido2016compressed, hansen2012retrospective, kellman2009high, jung2009k, gamper2008compressed} and have shown good reconstruction quality under high acceleration. The application of CS to RT-cine, however, is limited  by the slow reconstruction speed due to iterative algorithms and tricky hyperparameter tuning. Deep learning based methods, on the other hand, can potentially achieve much faster reconstruction speed and thus have drawn many attentions in recent years \citep{schlemper2017deep, hammernik2018learning, han2019k, hauptmann2019real, jin2019time}. Qin et al. \citep{qin2018convolutional} developed a convolutional recurrent neural network for dynamic CMR. While these studies show promising results, there are several limitations: First, simulated undersampled data from retro-cine rather than real RT-cine data are used for evaluation. Second, the acceleration rates are usually lower than 10x. Third, algorithms are tested on synthesized single-coil data rather than multi-coil data. Fourth, the reconstruction quality evaluations lack clinical assessment.

In this study, we propose a residual convolutional recurrent neural network (Res-CRNN) for RT-cine MRI. Our deep learning model is evaluated on highly accelerated (12x) multi-coil RT-cine data and compared to CS. To the best of our knowledge, this is the first work applying the deep learning approach to real Cartesian RT-cine MRI and it is evaluated by radiologists.


\section{Methods}
Since it is almost impossible to obtain ground truth (i.e., fully sampled data) for RT-cine, we adopt the strategy of training the deep learning model on retro-cine data. Then the trained model is applied to the acquired undersampled multi-coil RT-cine data directly from scanners for image reconstruction.  

\textbf{Data collection}: We collected retro-cine data of 51 patients (total 343 slices) with a bSSFP sequence on a clinical 3T scanner (uMR 790 United Imaging Healthcare, Shanghai, China) with the approval of local IRB. We also collected RT-cine data of 27 slices from two patients, which were acquired using a bSSFP sequence using variable Latin Hypercube undersampling \citep{Lyu2019toward}. Imaging parameters include: imaging matrix: 192 x 180, TR/TE = 2.8/1.3 ms, spatial resolution = $1.82\times1.82$ $mm^2$, and temporal resolution = 34 ms and 42 ms for retro-cine and RT-cine, respectively. Both retro- and RT-cine data were acquired using phased-arrayed coils.

\begin{figure}[htbp]
\centering
\includegraphics[width=\linewidth]{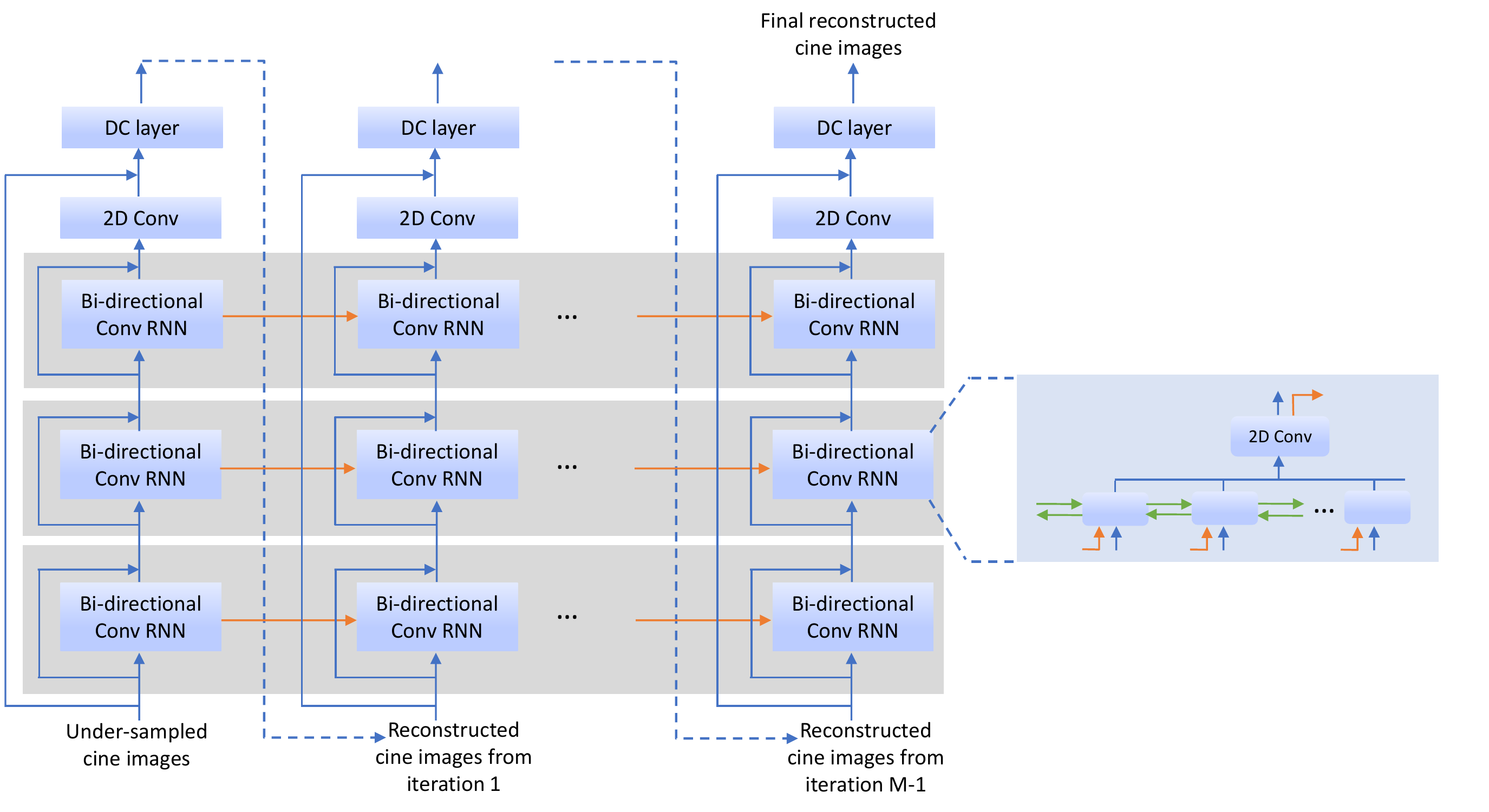}
\caption{Res-CRNN for RT-cine MRI reconstruction. The bi-directional ConvRNN layers are used to model dynamic information. Two levels of residual connections promote the network to learn high-frequency details. To reduce the GPU memory consumption and speed up reconstruction, one extra 2DConv layer is added within each bi-directional ConvRNN layer to reduce the number of feature maps in hidden states (red arrows). Reconstructed images from each intermediate iteration are then fed into the next iteration as input. The network includes five iterations. The network is fully convolutional and thus can take input with various image sizes and cardiac phases.}
\label{fig:network}
\end{figure}

\textbf{Res-CRNN}: Figure \ref{fig:network} depicts the proposed Res-CRNN. The network includes: three bi-directional convolutional RNN layers to model dynamic information \citep{qin2018convolutional}, the data consistency layer \citep{schlemper2017deep},  as well as two levels of residual connections, which promote the network to learn high-frequency details.  To reduce the GPU memory consumption and speed up reconstruction, one extra 2DConv layer is added within each bi-directional ConvRNN layer to reduce the number of feature maps in hidden states. All Conv kernels are $3\times3$ and the filter numbers are k=48 for bi-directional ConvRNN layers and k=2 for other 2DConv layers. 

\textbf{Model training}: For model training, the retro-cine data were retrospectively undersampled using the same 12x acceleration sampling mask as in RT-cine and then fed into the neural network as input. MSE and SSIM \citep{zhao2015loss} were used as the training loss. Images from each coil were reconstructed independently and then combined using the root sum of square method. 

\textbf{Evaluation}: For comparison, the same RT-cine data was reconstructed by CS using BART \citep{tamir2016generalized}. Temporal total variation and spatial wavelets were adopted and ADMM was used for optimization with maximum iterations of 100. CS hyperparameters were heuristically optimized on one representative data. Coil sensitivity maps were calculated using ESPIRiT \citep{uecker2014espirit} from temporally averaged data. Res-CRNN and CS were evaluated on a workstation (Intel Xeon E5-2630 CPU, 256G memory and Nvidia Tesla V100 GPU).

The reconstruction results were independently evaluated by four experienced cardiologists based on four radiologists (Figure 3), where a score in the 1-5 scale (1 as unacceptable and 5 as perfect) was assigned. The pair of reconstruction videos from CS and Res-CRNN were presented side-by-side in a randomized and blind fashion to radiologists.  Statistical significance was calculated by repeated measures ANOVA.  


\begin{figure}[htbp]
\centering
\includegraphics[width=\linewidth]{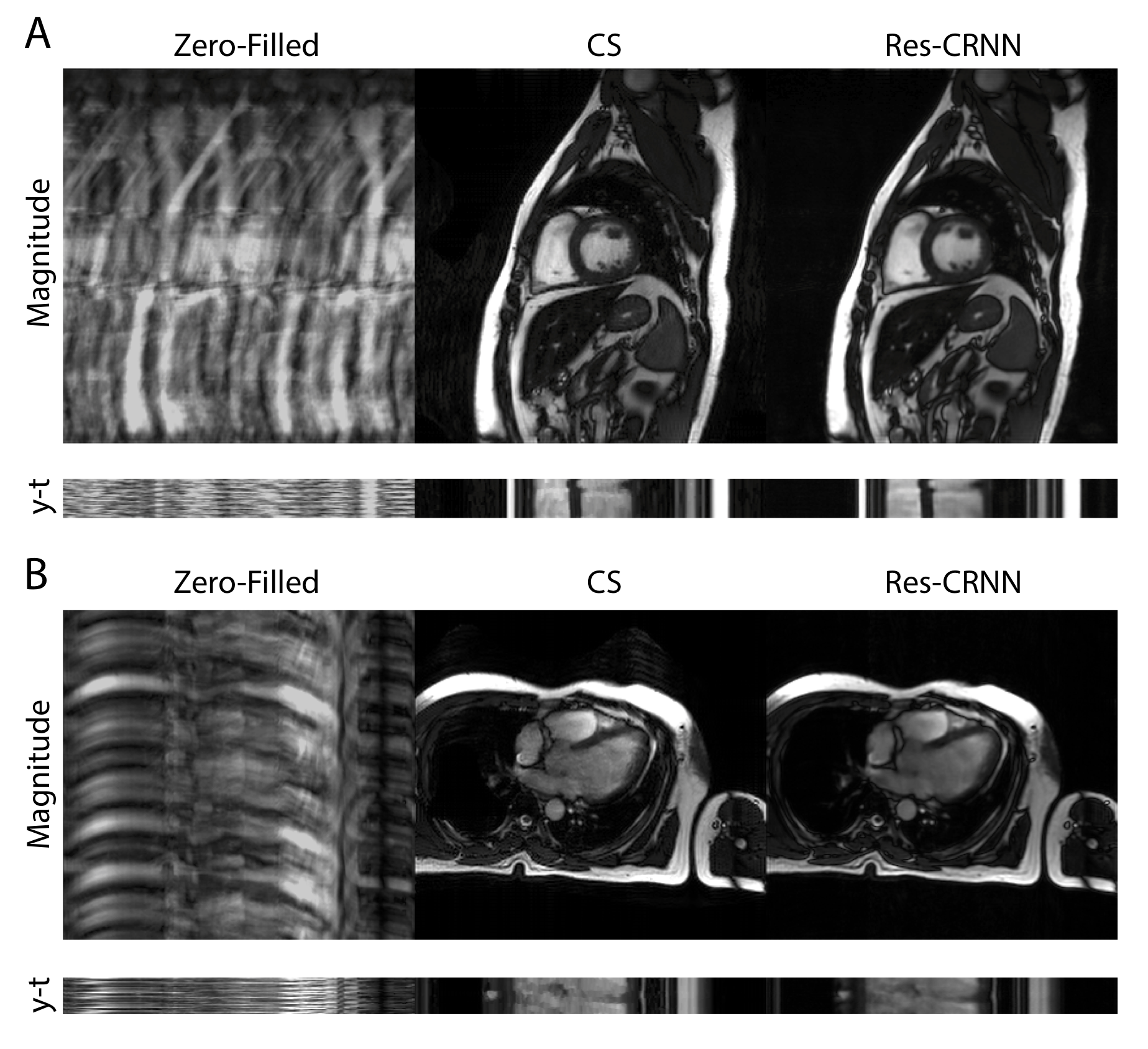}
\caption{Examples of RT-cine reconstruction of (A) short-axis cine images (B) long-axis cine images by zero-filled iFFT, compressed sensing (CS) and Res-CRNN at 12x acceleration rate. Note the aliasing artifacts along the phase-encoding direction in the blood pool regions as well as in the background of CS reconstruction (zoom in for more details). }
\label{fig:result_image}
\end{figure}

\begin{figure}[htbp]
\centering
\includegraphics[scale=0.95]{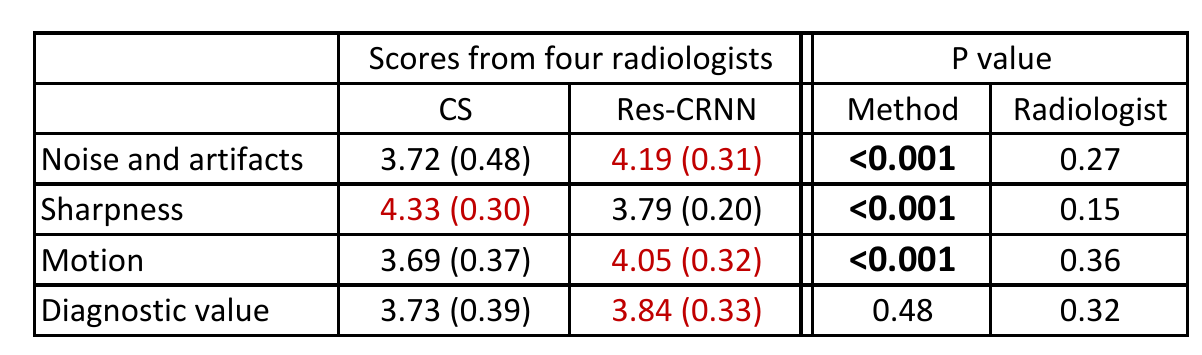}
\caption{Evaluation of RT-cine reconstruction. A total of 27 pairs of CS and Res-CRNN reconstructed RT-cine videos were independently evaluated by four radiologists in a randomized and blind fashion. The mean (standard deviation) of the scores are shown. Res-CRNN has significantly better scores for noise and artifacts as well as motion but lower scores for sharpness than CS (better scores are in red). There is no significant difference in diagnostic value between CS and Res-CRNN.}
\label{fig:result_table}
\end{figure}

\begin{figure}[htbp]
\centering
\includegraphics[scale=1.0]{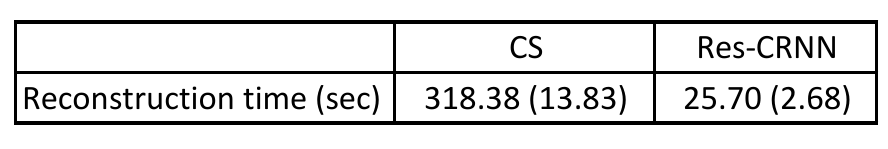}
\caption{Reconstruction time of CS and Res-CRNN. Time is shown as mean (standard deviation) in seconds. Res-CRNN is significantly faster than CS.}
\label{fig:result_time}
\end{figure}

\section{Results}
Figure \ref{fig:result_image} shows examples of RT-cine reconstruction results. Res-CRNN achieves less noise and aliasing artifacts than CS. Figure \ref{fig:result_table} shows the evaluation from four radiologists. Res-CRNN reconstruction receives significantly better scores for noise and artifacts as well as motion, but lower scores for sharpness than CS (all P$<$0.001). There is no significant difference in diagnostic value between CS and Res-CRNN (P$=$0.48). No significant difference among the four radiologists was observed (all P$>$0.1). Reconstruction speed of Res-CRNN is more than ten times faster than CS (318.3s vs. 25.7s on average).


\section{Discussion and Conclusion}
Unlike other studies using synthetic data from retro-cine \citep{qin2018convolutional, schlemper2017deep,Cheng2019dlespirit} or processed coil-combined data from RT-cine \citep{hauptmann2019real}, we proposed a deep learning model, Res-CRNN, for real multi-coil RT-cine MRI reconstruction. The deep learning method presents superior image quality over CS and is favored by the radiologists in noise/artifacts reduction and motion depiction. Moreover, the reconstruction speed of the deep learning method is significantly faster than CS, making it suitable for daily clinical usage. With implementation optimization, the reconstruction speed of Res-CRNN can be further improved. 


\bibliographystyle{abbrvnat}

\bibliography{references}
\end{document}